\documentclass[
preprint,
 amsmath,amssymb,
 aps,
]{revtex4-2}

\usepackage{graphicx}% Include figure files
\usepackage{dcolumn}% Align table columns on decimal point
\usepackage{bm}% bold math
\usepackage{float}
\usepackage{color}

\pdfoutput=1
\begin{document}

%\preprint{APS/123-QED}

\title{Mass Determination of Supermassive Black Holes Governing Evolution of Radio Emitters}

%%\thanks{A footnote to the article title}%

\author{Kimitake Hayasaki}
\email{kimi@chungbuk.ac.kr}
\affiliation{Department of Space Science and Astronomy, Chungbuk National University, Cheongju 361-763, Korea
\\
%\affiliation{
Department of Physical Sciences, Aoyama Gakuin University, Sagamihara 252-5258, Japan
}
%
%\altaffiliation[Also at ]{Department of Physical Sciences, Aoyama Gakuin University,  Sagamihara 252-5258, Japan}
\author{Ryo Yamazaki}%
\email{ryo@phys.aoyama.ac.jp}


\affiliation{
Department of Physical Sciences, Aoyama Gakuin University, Sagamihara 252-5258, Japan
\\
%\affiliation{
Institute of Laser Engineering, Osaka University, 2-6 Yamadaoka, Suita, Osaka 565-0871, Japan
}%
%\affiliation{%
% Authors' institution and/or address}%


\date{\today}% It is always \today, today,

\begin{abstract}
Tidal disruption events (TDEs) involving supermassive black holes (SMBHs) often exhibit radio emission, yet its physical origin remains uncertain, especially in non-jetted cases. In this Letter, we formulate a general dynamical framework for a radio-emitting shell driven by disk winds and expanding through a power-law ambient medium under the influence of SMBH gravity. We derive and classify power-law-in-time solutions to the governing equations in the adiabatic regime. In particular, a universal $t^{2/3}$ scaling emerges naturally when gravitational energy dominates or is comparable to thermal energy, irrespective of the ambient density profile, whereas the classical Sedov–Taylor solution is recovered when gravity is negligible. Our analysis reveals that, in regimes where SMBH gravity governs the shell expansion, the SMBH mass can be inferred from radio observations of the shell. This approach is independent of and complementary to conventional mass estimators, with direct implications for interpreting radio-emitting TDEs and probing SMBH demographics. Our formalism further predicts that 10–100\,GHz monitoring with existing and planned facilities can yield SMBH masses within months of disruption, providing a time-domain analogue to reverberation mapping.
\end{abstract}

                        %displa desired
\maketitle

%\tableofcontents

%
%%%%%%%%%%%%%%%%%%%%%%
\section{Introduction}
\label{sec:intro}
%%%%%%%%%%%%%%%%%%%%%%
%

Cosmic transients---short-lived astrophysical events with dramatic brightness changes---offer powerful diagnostics of extreme physical conditions in the universe. These phenomena include supernovae, gamma-ray bursts (GRBs), kilonovae, and tidal disruption events (TDEs), each linked to highly energetic processes. A TDE occurs when a star approaches a supermassive black hole (SMBH) closely enough to be torn apart by tidal forces \citep{hills_possible_1975, rees_tidal_1988}. The fallback of stellar debris forms an accretion disk, likely via relativistic apsidal precession, and produces a luminous flare detectable across multiple wavelengths \citep{rees_tidal_1988}. TDEs thus offer a unique opportunity to probe SMBH demographics and accretion physics in otherwise quiescent galactic nuclei \citep{stone_rates_2020}.

A subset of TDEs exhibit radio emission \citep{alexander_radio_2020}, generally attributed to synchrotron radiation from shocks involving relativistic electrons 
(e.g., Longair~2011~\cite{longair_high_2011}). 
While jetted TDEs such as Sw~J1644+57 \citep{bloom_possible_2011} are powered by 
relativistic jets, in which the dynamics of radio-emitting regions are typically dominated by ongoing energy injection from relativistic jets, and gravitational deceleration becomes subdominant, the majority of events show no such jet signatures. Hereafter we refer to these jet–less events as non-jetted TDEs. This category overlaps with what the literature often calls 
\emph{radio-quiet} TDEs when their peak radio luminosity satisfies 
$\nu L_\nu\lesssim10^{40}\,\mathrm{erg\,s^{-1}}$ 
\citep[][and references therein]{alexander_radio_2020}. 
Non-jetted TDEs nevertheless can produce mild radio outflows through shocks driven by 
disk winds, stream–stream collisions, or unbound ejecta 
\citep{jiang_prompt_2016,guillochon_unbound_2016,krolik_asassn-14li_2016,2023ApJ...954....5H}. 
However, the physical origin and time evolution of these radio-emitting regions remain poorly understood.

Although large-scale active galactic nucleus (AGN) winds and X-ray binary (XRB) disk outflows are highly anisotropic, their earliest over‑pressured phases can be quasi‑spherical before collimation develops (e.g., \cite{King2015,Faucher2012}). Such a quasi‑spherical geometry of the radio‑emitting region has also been inferred for non‑jetted TDEs, such as AT2019dsg, whose early radio follow‑up indicated a broad, quasi‑isotropic outflow \citep{Cendes2021,stein_tidal_2021}.
Observations show that only a few percent of TDEs produce powerful relativistic jets \cite{alexander_radio_2020,2025arXiv250605476L}, comparable to the $\sim8\%$ radio-loud fraction of optically selected quasars \cite{Ivezic2002}. The vast majority of both TDEs and AGNs are therefore classified as radio‑quiet, typically lacking persistent, large‑scale relativistic jets. Non‑jetted TDEs thus provide clean laboratories for studying a short‑lived, quasi‑spherical wind phase without jet-driven complications. By modeling the transition from gravity-dominated to energy-conserving expansion in non-jetted TDEs, we seek to clarify their connection to other outflow-driven phenomena, including AGN winds and stellar explosions \cite{MetzgerStone2016}.

\begin{figure}[H]
\centering
\includegraphics[width=0.65\columnwidth]
{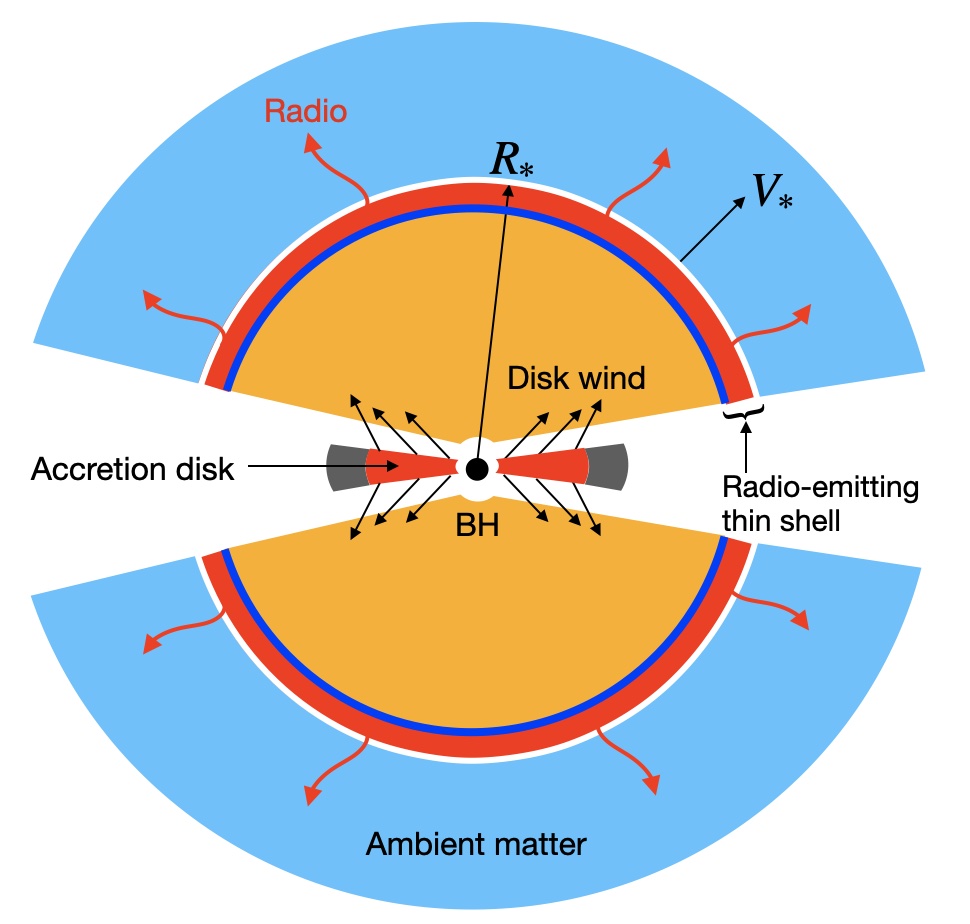}
\caption{
Schematic illustration of our model for a non-jetted TDE, showing the early-time evolution of a quasi-spherical, radio-emitting thin shell driven by disk winds and decelerated by the gravity of the central SMBH. This figure shows a meridional cross-section of the three-dimensional structure formed by the interaction between the disk wind and ambient matter surrounding the SMBH. The accretion disk (red part) forms from the fallback debris (gray part) and launches a quasi-isotropic wind (orange region). This wind sweeps up ambient matter (light blue), forming a geometrically thin shell (red). The shell expands outward and produces synchrotron radio emission (red wavy arrows). The shell’s radius $R_\ast$ and expansion velocity $V_\ast$ are typically inferred from broadband radio spectra through equipartition modelling (see Supplementary Table S1 \cite{supp} for data and references). Gravitational deceleration leads to a characteristic scaling relation, $M_{\rm bh} \propto R_\ast V_\ast^2$ (see equation~\ref{eq:bhmass}), enabling an independent method for SMBH mass estimation. The red region marks the shocked shell, bounded on the inside by the reverse shock (blue line) and on the outside by the forward shock (white line), so that the radial order is blue–red–white. Due to mixing across the contact discontinuity, the shocked wind and shocked ambient gas are treated together as a single shocked shell region. This model applies to non-jetted TDEs during the early, nearly spherical phase, in which the shell expansion asymptotically follows a power-law behaviour driven by SMBH gravity.
Relativistic jets, if present, are excluded from this scenario.
}
\label{fig:fig1}
\end{figure}

In non-jetted TDEs, the gravitational potential of the SMBH governs the evolution of the radio-emitting material, fundamentally altering its dynamics compared to classical shock models. Existing models developed in the context of supernovae and GRBs often assume adiabatic shock expansion, typically described by the Sedov-Taylor solution in the non-relativistic regime \citep{1950RSPSA.201..159T,1959sdmm.book.....S} and the Blandford-McKee solution in the relativistic regime \cite{1976PhFl...19.1130B}. However, these solutions appear when the expanding shell is dominated by thermal energy, depend on the ambient density profile, and neglect the gravitational influence of compact objects. A simple order-of-magnitude estimate yields the ratio of gravitational to thermal energy as $GM/(rv_{\rm s}^2) \sim 5 \times 10^{-9} (M/M_\odot)(T/10^8\,{\rm K})^{-1}(r/1\,{\rm pc})^{-1}$, where $G$ is the gravitational constant, $M$ is the mass of a compact object, $r$ is a radial distance from the central compact object, $v_{\rm s}$ is the shock velocity, and $T$ is the post-shock temperature corresponding to $v_{\rm s}$. The ratio remains much smaller than unity at large distances for a neutron star or a stellar-mass black hole. This suggests that gravitational effects are negligible for supernovae and GRBs. In contrast, for SMBHs, this ratio approaches or exceeds unity, suggesting that conventional solutions are no longer applicable to TDEs. Understanding the evolution of an expanding shell under SMBH gravity is therefore essential for interpreting TDE radio observations and constraining the underlying physics. A schematic overview of the shell geometry and the associated physical regions is presented in Figure~\ref{fig:fig1}.

In this Letter, we present a one-dimensional model for the expansion of a radio-emitting shell in non-jetted TDEs, where the mass injection rate and ambient density follow power-law profiles. Our analysis reveals a universal power-law scaling for the shell expansion that remains independent of the ambient density slope---unlike conventional solutions. A dimensional analysis incorporating the gravitational constant and mass implies that the radius scales as the two-thirds power of time: $r\propto(GM)^{1/3}t^{2/3}$. 
This scaling arises naturally from our model, offering a direct prediction for radio observations and an independent method to infer black hole masses, distinct from traditional approaches such as reverberation mapping (e.g.,~\cite{2004AN....325..248P}). We formulate the fundamental equations governing the radio-emitting thin shell in Sec. II, and present analytical solutions to them, including one incorporating gravitational effects, in Sec. III. In Sec. IV, we compare our mass estimates with existing methods and discuss the validity of our theoretical framework in light of these comparisons. Sec.~V summarizes our conclusions.

%
%%%%%%%%%%%%%%%%%%%%%%%%%%%%%%%
\section{Basic equations for an expanding thin shell}
\label{sec:basiceqs}
%%%%%%%%%%%%%%%%%%%%%%%%%%%%%%%
%
We investigate the dynamics of an expanding thin shell driven by TDE disk winds, incorporating SMBH gravity, mass accretion from the ejecta, ambient drag, and thermal pressure. Extending a previous study \citep{2023ApJ...954....5H} that considered only momentum conservation, we systematically derive governing equations based on momentum and energy conservation. Our shell model provides a unified framework describing the shell evolution from formation onward, capturing, e.g., the free-expansion and ejecta-dominated phases before asymptotically transitioning to a $t^{2/3}$ scaling under SMBH gravity. In contrast to self-similar solutions for wind or accretion, which show $t^{2/3}$ evolution due to the dimension analysis from the outset \citep{1974Ap&SS..26..183S,1994ApJ...429..759P}, our model naturally recovers this asymptotic behavior while suggesting that more complex shell dynamics is incorporated in earlier phases. Moreover, in the absence of gravitational effects, our model reproduces multiple asymptotic solutions, including the Sedov-Taylor expansion when thermal pressure dominates, demonstrating its versatility in describing a broad class of shell dynamics. Supplemental Material describes the derivation of the detailed formulae in this section \cite[See Supplemental Material at ][ for the detailed derivation of the basic equations and their solutions.]{supp}. Hereafter, an overdot denotes a derivative with respect to time, as in $\dot{Q} = dQ/dt$ for an arbitrary physical quantity $Q$.

%
%%%%%%%%%%%%%%%%%%%%%%%%%%%%%%
\subsection{Shell Mass}
%Expanding Shell Mass}
%Mass of the Expanding Shell}
%%%%%%%%%%%%%%%%%%%%%%%%%%%%%%
% 
A star is tidally disrupted by a SMBH, leading to the formation of an accretion disk at $t<t_0$. The disk subsequently drives an outflow as ejecta, which interacts with the ambient medium and results in the formation of an expanding thin shell at $t=t_0$ \cite{2023ApJ...954....5H}. We assume that the mass element of the shell moving under the SMBH gravity along the radial direction has an escape velocity as 
\begin{eqnarray}
\dot{r}=\sqrt{\frac{2GM_{\rm bh}}{r}},
\label{eq:rdot}
\end{eqnarray}
where $M_{\rm bh}$ is the SMBH mass. The initial shell velocity is then estimated to be $v_0=\sqrt{2GM_{\rm bh}/r_0}$, where $r_0$ is set to be the tidal disruption radius. We assume that the mass is ejected from the disk to be injected into the shell at the following rate:
\begin{eqnarray}
\dot{M}_{\rm inj} (t) &=& \dot{M}_0 \left( \frac{t}{t_0} \right)^{-n},
\label{eq:mdotinj}
\end{eqnarray}
where $\dot{M}_0$ is the initial mass ejection rate and $n=5/3$ is the power-law index associated with mass fallback rate onto the SMBH \citep{rees_tidal_1988}. In order to determine the ejecta mass profile $\rho_{\rm ej}(r,t)$, we use the fact that mass ejected from the disk at an earlier coordinate time $t'$ contributes to the shell only after a finite travel time under the gravitational potential of the SMBH. This one-to-one mapping ensures conservation of mass between the injection point at $r = r_0$ during the interval $\Delta t'$ and the radial segment from $r$ to $r + \Delta r$ at the current coordinate time $t$. The relation between the current coordinate time $t$ and its retarded time $t'$ therefore yields
\begin{equation}
    \dot{M}_{\rm inj}(t^\prime)\Delta{t^\prime}=4\pi{r}^2\Delta{r}\rho_{\rm ej}(r,t),
%    \nonumber
\label{eq:masscon} 
\end{equation}
where $t^\prime$ is the retarded time, which is given by
\begin{eqnarray}
t^\prime
=
t
+
\frac{2}{3}
\epsilon
%\frac{r_0}{v_0}
\left[
1
-
\left(\frac{r}{r_0}\right)^{3/2}
\right]
t_0
\label{eq:tret}
\end{eqnarray}
with a dimensionless velocity parameter $\epsilon=(r_0/t_0)/v_0$. Substituting the relation on $t$ and $t^\prime$: $\Delta{r}=\dot{r}\Delta{t^\prime}$ and equations~(\ref{eq:rdot}), 
(\ref{eq:mdotinj}), and (\ref{eq:tret}) into equation (\ref{eq:masscon}) yields the unshocked ejecta's mass density $\rho_{\rm ej}(r,t)$ as equation~(9) of Supplemental Material \cite{supp}. The volume integral gives the ejecta mass comprising the infinitesimally thin shell at radius $R(t)$ as
\begin{eqnarray}
M_{\rm ej}(t)
&=&
\int_{R(t)}^{r_{\rm max}}\,4\pi{r^2}\rho_{\rm ej}(r,t)\,dr,
\label{eq:mej}
\end{eqnarray}
where $r_{\rm max}$ is the farthest radius that the wind's mass element can reach, which is given by equation~(4) of Supplemental Material \cite{supp}.

Integrating the ambient mass density, which is assumed be the power-law distribution, $\rho_{\rm am}=\rho_{\rm am,0}(r/r_0)^{-s}$ with the power-law index $s$, from $r_0$ to $R(t)$, we obtain the shell mass, to which is contributed from the ambient matter, as
\begin{equation}
M_{\rm am}(t)
=
\int^{R(t)}_{r_0}\,4\pi{r^2}\rho_{\rm am}(r)\,dr,
\label{eq:mam}
\end{equation}
where $\eta=\rho_{\rm am,0}/\rho_{\rm ej,0}$ is defined as a density ratio parameter and $\rho_{\rm ej,0}=\epsilon M_0/(4\pi r_0^3)$ with $M_0\equiv\dot{M}_0t_0$. We introduce the initial shell mass as   
\begin{equation}
\Delta{m}=4\pi\delta\rho_{\rm ej,0}r_0^3,
\label{eq:inimass} 
\end{equation}
where $\delta$ is a parameter to decide the amount of the initial shell mass. Equations~(\ref{eq:mej}), (\ref{eq:mam}), and (\ref{eq:inimass}) yields the total mass of the thin shell at radius $R(t)$ as
\begin{eqnarray}
M(t)
&=&
M_{\rm ej}(t) + M_{\rm am}(t) + \Delta{m}.
\label{eq:totm}
\end{eqnarray}
The detailed expression for the shell mass is provided in equations~(18) of the Supplemental Material \cite{supp}.

%
%%%%%%%%%%%%%%%%%%%%%%%%%%%%%%%%%%
\subsection{Momentum Conservation}
%%%%%%%%%%%%%%%%%%%%%%%%%%%%%%%%%%
%
The principle of momentum conservation describes the change in momentum of the shell due to an impulse $F(t) \Delta{t}$ 
occurring between $t$ and $t + \Delta{t}$ as 
\begin{eqnarray}
F(t)\Delta{t}
&&
=M(t+\Delta{t})\dot{R}(t+\Delta{t})-M(t)\dot{R}(t),
\label{eq:momcon}
\end{eqnarray}
where $F(t)$ is any force acting on the shell. 
This force is the sum of the mass-ejection-driven force, gravitational force, and thermal pressure force:
\begin{eqnarray}
F(t)=F_{\rm ej}(t) + F_{\rm G}(t)  + F_{\rm th}(t),
\label{eq:forces}
\end{eqnarray}
where 
\begin{eqnarray}
F_{\rm ej}(t)
&=&
\dot{M}_{\rm ej}\dot{r}(R(t))
\label{eq:fej}
\end{eqnarray}
is the ram pressure force induced by the ejecta, 
\begin{eqnarray}
&&
F_{\rm G}(t)
=-\frac{GM_{\rm bh}M(t)}{R(t)^2}
\label{eq:fg}
\end{eqnarray}
is the gravitational force of the SMBH, and 
\begin{eqnarray}
F_{\rm th}(t)
&&
=
4\pi{R(t)^2}\chi{P}(t)
\label{eq:fth}
\end{eqnarray}
is the thermal pressure force acting on the shell from the outside of the shell with the thermal energy $E_{\rm T}$, respectively. Here the shell's thermal pressure is given by $P(t) = (\gamma - 1) E_{\rm T}(t) / \left(4\pi R(t)^2 \Delta R\right)$, where $\gamma = 5/3$ is the adiabatic index, $\chi$ is the ratio of thermal pressure inside and outside the shell, and $\xi$ denotes the ratio of the shell thickness $\Delta{R}$ to the shell size as $\xi=\Delta{R}/R(t)$ with a representative value of $\xi\approx0.1$ \citep{cendes_radio_2021}.

Neglecting terms beyond the second order in $\Delta t$ in the Taylor expansion of the right-hand side of equation~(\ref{eq:momcon}) in the limit $\Delta t \rightarrow 0$, and applying the force expressions given in equations~(\ref{eq:forces})-(\ref{eq:fth}), we obtain the equation of motion as
\begin{eqnarray}
M(t) \ddot{R}(t) 
&=& 
\dot{M}_{\rm ej} (t) \dot{r}(R(t)) - \dot{M}(t) \dot{R}(t) - \frac{G M_{\rm bh} M(t)}{R(t)^2} 
\nonumber \\
&+& \frac{\chi(\gamma - 1)}{\xi} \frac{E_T(t)}{R(t)}.
\label{eq:dpdt}
\end{eqnarray}
The detailed expressions for $\dot{M}_{\rm ej}(t)$ and $\dot{M}(t)$ are given in equations~(12) and (20) of the Supplemental Material \cite{supp}.

%
%%%%%%%%%%%%%%%%%%%%%%%%%%%%%%%%
\subsection{Energy Conservation}
%%%%%%%%%%%%%%%%%%%%%%%%%%%%%%%%
%
The conservation of total energy in the shell system, accounting for the ejecta, shell, and ambient matter before and after their interaction, is given by
\begin{eqnarray} 
E_{\rm K}(t) 
&+& 
E_{\rm T}(t) + \Phi(t) + E_{\rm C}(t) 
= 
E_{\rm K}(t + \Delta t) 
\nonumber \\
&+& 
E_{\rm T}(t + \Delta t) + \Phi(t + \Delta t) + E_{\rm C}(t + \Delta t),
\label{eq:enecon} 
\end{eqnarray} 
where 
\begin{eqnarray}
E_{\rm K}(t)
&=&
\frac{1}{2}\Delta{M}_{\rm ej}\dot{r}(R(t))^2 + \frac{1}{2}M(t)\dot{R}(t)^2
\label{eq:ekin}
\end{eqnarray}
represents the kinetic energy with the mass of the ejecta added due to collision with the shell during $\Delta{t}$, i.e., $\Delta{M}_{\rm ej}$, 
\begin{eqnarray} 
\Phi(t) = \int_{R(t)}^{\infty}\,F_{\rm G}(r)\,dr  
\label{eq:gpot} 
\end{eqnarray} 
is the gravitational potential energy, and $E_{\rm C}(t)$ is the energy lost due to radiative cooling. Here, we assume that the thermal energy, $E_{\rm T}(t)$, is stored only in the shell. In the limit $\Delta t \to 0$, neglecting terms of order $\mathcal{O}(\Delta t^2)$ and higher in the Taylor expansion of equation~(\ref{eq:enecon}) with equations (\ref{eq:ekin}) and (\ref{eq:gpot}) yields the following expression for the thermal energy evolution of the shell: 
\begin{eqnarray}
\frac{dE_{\rm T}(t)}{dt}
&=&
\frac{1}{2}\dot{M}_{\rm ej}(t)
\biggr(
\dot{r}(R(t))
-
\dot{R}(t)
\biggr)^2
+
\frac{1}{2}\dot{M}_{\rm am}(t)\dot{R}^2(t)
\nonumber \\
&+&
\frac{GM_{\rm bh}\dot{M}(t)}{R(t)}
-
\frac{\chi(\gamma-1)}{\xi}
\frac{\dot{R}(t)}{R(t)}
E_{\rm T}(t)
-
\dot{E}_{\rm C}(t),
\nonumber \\
\label{eq:thermalenergyevo}
\end{eqnarray}
where the first two terms correspond to kinetic energy contributions from ejecta and ambient mass, followed by gravitational and thermal pressure effects, with $\dot{E}_{\rm C}$ representing radiative losses. Potential cooling mechanisms include free-free emission and synchrotron radiation. Radio observations of non-jetted TDEs suggest that cooling due to synchrotron emission is insignificant \citep{alexander_radio_2020,2024ApJ...971..185C}, while thermal bremsstrahlung cooling is also negligible in the radio-emitting region (see Section IV of the Supplemental Material for details \cite{supp}). Therefore, in the following sections, we disregard the $\dot{E}_{\rm C}$ term.

Our formulation generalizes the previous treatment \citep{2023ApJ...954....5H} by incorporating comprehensive energy considerations, revealing how mass accretion, drag, and thermal pressure collectively influence shell dynamics. 
%These results 
Equations~(\ref{eq:dpdt}) and (\ref{eq:thermalenergyevo}) together with equations (\ref{eq:mej}), (\ref{eq:mam}), and (\ref{eq:totm}) provide a foundation for interpreting non-relativistic radio-emitting TDEs and can be extended to other astrophysical outflows. 
A complete summary of the governing equations for the time evolution of an expanding thin shell is provided in equations~(37)-(39) of Supplemental Material.

%
%%%%%%%%%%%%%%%%%%%%%%%%%%%%%%%%%%%%%%%%%%%
\section{Solutions for the basic equations}
%%%%%%%%%%%%%%%%%%%%%%%%%%%%%%%%%%%%%%%%%%%
%

Our interest lies in understanding the asymptotic behavior of an adiabatically expanding thin shell at large distances from the central SMBH, where the evolution can be directly compared with radio observations. 
To this end, we seek analytic solutions to the governing equations in the asymptotic regime, applicable to both thermal pressure-driven and momentum-driven expanding shells. 
Among these scenarios, we focus on the case in which thermal pressure and SMBH gravity jointly govern the shell dynamics. This solution enables an estimate of the SMBH mass based on the observed expansion profile.
Detailed derivations of the corresponding asymptotic solutions are provided in the Supplemental Material \cite{supp}.

%
%%%%%%%%%%%%%%%%%%%%%%%%%%%%%%%%%%%%%%%%%%%%%%%%%
\subsection{
%Power-Law Solution Under the Gravity
Solutions Incorporating Gravity
%Power-Law Solutions Incorporating Gravity
}
%%%%%%%%%%%%%%%%%%%%%%%%%%%%%%%%%%%%%%%%%%%%%%%%%
%

For sufficiently large radii, $ R(t) \gg r_0 $ (or $t\gg t_0$), the total mass of the shell asymptotically approaches a form dominated by the accumulated ambient material, i.e., $M(t)\approx M_{\rm am}(t)$ in equation~(\ref{eq:totm}). Consequently, equations~(\ref{eq:dpdt}) and (\ref{eq:thermalenergyevo}) are reduced to:
\begin{eqnarray}
M_{\rm am}(t)\ddot{R}(t)
&=&
-
\dot{M}_{\rm am}(t)\dot{R}(t)
-
\frac{GM_{\rm bh}M_{\rm am}(t)}{R(t)^2}
\nonumber \\
&+&
\frac{
\chi(\gamma-1)
}{\xi}
\frac{E_{\rm T}(t)}{R(t)},
\label{eq:dpdt-am}
\\
\dot{E}_{\rm T}(t)
&=&
\frac{1}{2}\dot{M}_{\rm am}(t)\dot{R}(t)^2
+
\frac{GM_{\rm bh}\dot{M}_{\rm am}(t)}{R(t)}
\nonumber \\
&-&
\frac{\chi(\gamma-1)}{\xi}
\frac{\dot{R}(t)}{R(t)}
E_{\rm T}(t).
\label{eq:dedt-th}
\end{eqnarray}

To find power-law solutions, we assume a power-law form:
\begin{eqnarray}
R(t) = R_*
\left(\frac{t}{t_*}\right)^m,
\label{eq:plform}
\end{eqnarray}  
where $R_*$ and $t_*$ are the normalization radius and time, respectively, and $m$ is the scaling exponent. 
Substituting equation~(\ref{eq:plform}) into equations~(\ref{eq:dpdt-am}) and (\ref{eq:dedt-th}) and equating the time-dependent terms on both sides yields:
\begin{eqnarray}
    m=\frac{2}{3}.
    \label{eq:m23}
\end{eqnarray}
Note that this index corresponds to the regime in which the gravitational force is dominant or even comparable to the thermal pressure force. With the value of $m$ fixed by equation~(\ref{eq:m23}), the black hole mass can be inferred by comparing the coefficients of the substituted equations:
\begin{eqnarray}
M_{\rm bh}
&=&
K
\frac{R_{*}V_*^2}{G},
\label{eq:bhmass}
\end{eqnarray}
where $V_* = m (R_* / t_*)=(2/3)R_\ast/t_\ast $ represents the shell velocity at $t=t_*$ and $K$ is the proportionality coefficient given by 
\begin{eqnarray}
K
&=&
\frac{1}{2}\left[
(\gamma-1)
-\frac{\xi}{\chi}
\right]^{-1}
\left[
\gamma-1
+
(5-2s)
\frac{\xi}{\chi}
\right].
\nonumber
\end{eqnarray}
This equation imposes the constraint $\xi/\chi < \gamma - 1$; for $\gamma = 5/3$ and $s < 5/2$, this yields $\xi/\chi < 2/3$. In the thin-shell limit, $\xi/\chi \ll 1$, the coefficient $K$ asymptotically approaches $1/2$, regardless of the values of $\gamma$ and $s$. Since both $R_\ast$ and $V_\ast$ can be inferred from radio observations 
(e.g., \cite{alexander_discovery_2016}; see also Table~S1 in the Supplemental Material~\cite{supp}), equation~(\ref{eq:bhmass}) provides a direct means of estimating the black hole mass from observed shell expansion.

%
%%%%%%%%%%%%%%%%%%%%%%%%%%%%%%%%%%%
\subsection{Sedov-Taylor Solutions}
%%%%%%%%%%%%%%%%%%%%%%%%%%%%%%%%%%%
%

Equating the second and third terms on the right-hand side of equation~(\ref{eq:dpdt-am}), we get the critical mass to decide whether thermal pressure force is greater than the gravitational force in equation~(\ref{eq:dpdt-am}) as
\begin{eqnarray}
M_{\rm c} 
&=& 
\frac{\chi(\gamma-1)}{\xi}\frac{R_{\ast}}{GM_{\rm am,\ast}}E_{\rm T,\ast} 
\nonumber \\
&\sim&
%6.2
6
\times10^{7}\,M_\odot
\,
\left(\frac{R_\ast}{10^{16}\,{\rm cm}}\right)
\left(\frac{T_{p,\ast}}{10^{10}\,{\rm K}}\right),
\label{eq:mcrit} 
\end{eqnarray}
where $M_{\rm am,\ast}$, $E_{\rm T,\ast}$, and $T_{p,\ast}$ are the ambient matter mass, the thermal energy, and the proton temperature of the shell at $R_\ast$, respectively (see equations~(50), (52), and (53) of Supplemental Material for the detailed definitions of $M_{\rm am,\ast}$, $T_{p,\ast}$, and $E_{\rm T,\ast}$, respectively \cite{supp}).

For $M_{\rm bh}\ll M_{\rm c}$, the governing equations, i.e., 
equations~(\ref{eq:dpdt-am}) and (\ref{eq:dedt-th}) simplify to:
\begin{eqnarray}
M_{\rm am}(t)\ddot{R}(t)
&=&
-
\dot{M}_{\rm am}(t)\dot{R}(t)
+
\frac{
\chi(\gamma-1)
}{\xi}
\frac{E_{\rm T}(t)}{R(t)},
\label{eq:dpdt-am2}
\nonumber
\\
\dot{E}_{\rm T}(t)
&=&
\frac{1}{2}\dot{M}_{\rm am}(t)\dot{R}^2(t)
-\frac{\chi(\gamma-1)}{\xi}\frac{\dot{R}(t)}{R(t)}E_{\rm T}(t).
\nonumber 
\label{eq:dedt-th2}
\end{eqnarray}
Using a power-law ansatz, we find that the solution follows $R(t)\propto t^{2/(5 - s)}$, i.e., a Sedov-Taylor expansion when the gravitational force is negligible compared to the thermal pressure force, i.e., $ M_{\rm bh} \ll M_{\rm c} $.

%
%%%%%%%%%%%%%%%%%%%%%%%%%%%%%%%%%%%%%%%%%%%%%%%%
\subsection{Momentum-Driven Snowplow Solutions}
%%%%%%%%%%%%%%%%%%%%%%%%%%%%%%%%%%%%%%%%%%%%%%%%
%

If the kinetic energy of the ambient medium dominates over both the gravitational potential and thermal energy at $R_* \gg r_0$, equations~(\ref{eq:dpdt-am}) and (\ref{eq:dedt-th}) simplify to
\begin{eqnarray} 
M_{\rm am}(t)\ddot{R}(t) &=& -\dot{M}_{\rm am}(t)\dot{R}(t), 
\nonumber \\ 
\dot{E}_{\rm T}(t) &=& \frac{1}{2}\dot{M}_{\rm am}(t)\dot{R}(t)^2. 
\nonumber 
\end{eqnarray} 
Substituting the power-law form given in equation~(\ref{eq:plform}) into the above yields a power-law of time solution with $R(t)\propto{t}^{1/(4 - s)}$ \cite{2023ApJ...954....5H}. This result characterizes the momentum-driven regime, where both the gravitational influence of the SMBH and the thermal pressure become dynamically negligible.

%
%%%%%%%%%%%%%%%%%%%%
\section{Discussion}
%%%%%%%%%%%%%%%%%%%%
%

Among the growing sample of radio‐detected TDEs, AT2019dsg stands out because its radio shell was monitored from early to late times \citep{cendes_radio_2021,2024ApJ...971..185C}. Using radio observations of AT2019dsg \citep{cendes_radio_2021} in the regime relevant to our model, we estimate the SMBH mass from equation~(\ref{eq:bhmass}) as $M_{\rm bh} \sim 1 \times 10^8M_\odot (K/0.5)(R_\ast/10^{16.3}\,{\rm cm})^3 (t_\ast/130\,{\rm days})^{-2}$.
This estimate exceeds by roughly an order of magnitude the previously observed values $M_\bullet$, which were obtained by X-ray observations as
$M_{\bullet} \sim 5 \times 10^6M_\odot$ \citep{2021MNRAS.504..792C} and 
by the $M$–$\sigma$ relation as
$M_{\bullet} \sim 3 \times 10^7M_\odot$ \citep{stein_tidal_2021}. 
However, this discrepancy does not necessarily indicate a flaw in our theoretical framework, as discussed below.

Using equation~(\ref{eq:mcrit}), the condition $M_{\rm c} = M_{\bullet}$ defines a critical dimensionless parameter that serves as an indicator for determining whether the $m=2/3$ solution or the Sedov–Taylor solution is applicable: 
\begin{eqnarray}
\mathcal{D}_{\rm c}
\equiv
\frac{M_\bullet}{M_{\rm c}}
&=&
\frac{(\gamma+1)^2}{2\kappa(\gamma-1)}
\frac{G M_{\bullet}}{R_\ast v_{\rm s}^2}
%\nonumber 
\label{eq:dc}
\\
&\sim &
1
\kappa^{-1}
\left(\frac{R_\ast}{10^{16}\,{\rm cm}}\right)^{-1} 
\left(\frac{M_{\bullet}}{10^{7.5}M_\odot}\right)
\left(\frac{v_{\rm s}}{0.05\,c}\right)^{-2}
,
\nonumber
\end{eqnarray}
where $\kappa = (3/2)(\gamma - 1)(3 - s) \approx 1$ is adopted and equations~(58) of Supplemental Material were used for the derivation \cite{supp}. When $\mathcal{D}_{\rm c}\ll1$, the Sedov–Taylor solution governs the shell dynamics and equation~(\ref{eq:bhmass}) is not applicable. In contrast, the gravity-dominated $m=2/3$ solution is relevant only when $\mathcal{D}_{\rm c}>1$; values in the range $0.5\lesssim\mathcal{D}_{\rm c}\lesssim1$ represent a transition regime where both scalings coexist.
For TDEs involving SMBHs with $M_{\bullet}\lesssim10^{7}M_\odot$ and $v_{\rm s}\sim0.1c$, we find that $\mathcal{D}_{\rm c}\ll1$ at $R_\ast\gtrsim10^{16}$ cm, indicating that the Sedov–Taylor regime is realized. If instead $\mathcal{D}_{\rm c}>1$, the $m=2/3$ solution holds and equation~(\ref{eq:bhmass}) can be used to estimate $M_{\bullet}$ reliably.

Table~S1 compiles the shell radii and expansion velocities $(R_\ast, V_\ast)$  for six non-jetted TDEs, all inferred by fitting their multi-frequency radio spectra with synchrotron self-absorption equipartition models \cite{supp}. Applying $v_{\rm s}=V_\ast$ to equation~(\ref{eq:dc}) we estimate $\mathcal{D}_{\rm c}$ for each event. Four events (ASASSN-14li, AT2019dsg, AT2019azh, AT2020opy) have $\mathcal{D}_{\rm c}\lesssim0.06$, firmly placing them in the deep Sedov–Taylor regime. AT2020vwl and eRASStJ2344 yield intermediate values, $\mathcal{D}_{\rm c}=0.18$ and $0.84$, respectively. The latter lies in the transition regime and is closest in our sample to the gravity-dominated boundary at $\mathcal{D}_{\rm c}=1$. 
Equation~(\ref{eq:bhmass}) predicts $M_{\rm bh}=8.9\times10^{7}M_\odot$ for eRASSt~J2344, which is approximately $40\%$ larger than the value obtained with the existing method (e.g. \cite{2025ApJ...981..122G}). 
The span $0.002\le\mathcal{D}_{\rm c}\le0.84$ confirms that our formalism covers the full range from deep Sedov–Taylor expansion up to the onset of gravity domination and explains why equation~(\ref{eq:bhmass}) overestimates $M_{\bullet}$ for AT2019dsg. Discovering events with $\mathcal{D}_{\rm c}>1$ would provide a clean test-bed for the $m=2/3$ regime and enable direct SMBH-mass measurements from radio-measured shell dynamics.

Because $\mathcal{D}_{\rm c}$ depends only on the outflow radius and expansion speed, it can be estimated for any transient accretion episode once these two quantities are observationally constrained. 
This makes $\mathcal{D}_{\rm c}$ a promising scale-free diagnostic for comparing wind dynamics across black-hole systems of all masses, from XRBs to AGNs. It is particularly useful for systems that exhibit transient quasi-spherical outflows. If the relevant expansion regime (e.g., gravity- or energy-dominated) can be identified from observational trends, $\mathcal{D}_{\rm c}$ may also provide a purely kinematic means to infer the black-hole mass in future radio monitoring.

Two key implications follow from this $\mathcal{D}_{\rm c}$-based criterion. First, for TDEs associated with relatively low-mass SMBHs ($M_{\bullet} \lesssim 10^{7}M_\odot$), which dominate current observations, equation~(\ref{eq:bhmass}) becomes applicable only at smaller emission radii. According to standard equipartition analysis of radio observations, the peak frequency of the synchrotron spectrum scales inversely with the source size. A smaller $R_\ast$ implies that the spectral peak lies above the 1–10 GHz band, where the Very Large Array (VLA) operates most effectively. We therefore advocate high-cadence monitoring with the Atacama Large Millimeter/submillimeter Array (ALMA), which offers superior sensitivity in the 10–100 GHz range. Such observations will be essential for testing and constraining our model. Future facilities such as the next-generation VLA (ngVLA) will extend this capability with even higher sensitivity and broader frequency coverage.

Second, for radio-emitting regions with $R_\ast \gtrsim 10^{16}$cm, the $m = 2/3$ regime can be realized in TDEs involving more massive SMBHs ($M_{\bullet}\gtrsim10^8M_\odot$), rendering equation~(22) applicable. The maximum SMBH mass capable of producing a TDE, known as the Hills mass \cite{hills_possible_1975}, can reach $\sim 1\times10^9\,M_\odot$ for rapidly spinning black holes \citep{2024MNRAS.527.6233M}. Our mass estimation method is thus particularly suited for non-jetted, radio-emitting TDEs involving such high-mass, spinning SMBHs.

%
%%%%%%%%%%%%%%%%%%%%
\section{Conclusion}
%%%%%%%%%%%%%%%%%%%%
%

In this study, we derived the governing equations for the time evolution of a non-relativistically expanding thin shell in the gravitational field of a black hole, based on the conservation of momentum and energy. Applying asymptotic approximation at large radii, we identified power-law solutions for the shell expansion in the adiabatic regime. In particular, we found that when gravitational energy dominates over thermal energy, the expansion follows a universal scaling law with index $2/3$, independent of the ambient density profile. We further demonstrated that our framework naturally recovers the Sedov-Taylor solution in the limit where black hole gravity is negligible, and transitions to the momentum-driven snowplow solution when both gravity and thermal pressure can be ignored. Importantly, we found that in regimes where black hole gravity governs the shell dynamics, the black hole mass can be inferred from radio observations of the expanding shell. This provides an independent and complementary method for estimating black hole masses, particularly in systems where conventional techniques are challenging to apply.

%
%%%%%%%%%%%%%%%%%%%%%%%
\begin{acknowledgments}  
K.H. has been supported by the Basic Science Research Program through the National Research Foundation of Korea (NRF) funded by the Ministry of Education (2016R1A5A1013277 and 2020R1A2C1007219). This work was also supported in part by the National Science Foundation under Grant No. NSF PHY-1748958. R.Y. has been supported by JSPS KAKENHI 
(Grant Nos. 22H00119, 23K22522,  23K25907, 23H04899, and 24K00605).
\end{acknowledgments}
%%%%%%%%%%%%%%%%%%%%%
%

%\appendix
%\section{Appendixes}

\bibliographystyle{apsrev4-2}
\bibliography{arXiv}

\end{document}